\documentclass
[nofootinbib,titlepage,prl,tightenlines,preprint,onecolumn,floats]{revtex4}%
\usepackage{amsfonts}
\usepackage{amsmath}
\usepackage{amssymb}
\usepackage{graphicx}%
\begin{document}
\title{ Topology and Computational Performance of Attractor Neural Networks}
\author{Patrick N. McGraw and Michael Menzinger}
\email{pmcgraw@chem.utoronto.ca}
\affiliation{Department of Chemistry, University of Toronto, Toronto, ON M5S 3H6, Canada }
\date{\bigskip\bigskip\bigskip March 31, 2003\bigskip}

\begin{abstract}
To explore the relation between network structure and function, we studied the
computational performance of Hopfield-type attractor neural nets with regular
lattice, random, small-world and scale-free topologies. \ The random
configuration is the most efficient for storage and retrieval of patterns by
the network as a whole. However, in the scale-free case retrieval errors are
not distributed uniformly: the portion of a pattern encoded by the subset of
highly connected nodes is more robust and efficiently recognized than the rest
of the pattern. \ The scale-free network thus achieves a very strong partial
recognition. \ Implications for brain function and social dynamics are suggestive.

\end{abstract}
\maketitle

While intense research activity is centered on structural and topological
properties of social, biological and technological networks\cite{netreviews},
the consequences of network structure for the dynamics of cooperative
processes have been addressed to a lesser extent. \ Topology is known to
affect the ordering and disordering of the Ising model\cite{Pekalski}-\cite{G.
Bianconi} and the synchronization of coupled oscillators \cite{Barahona}%
\cite{Manrubia}. \ Another area of burning interest is the relation between
structure and function in the organization of brains. \cite{Braitenburg}%
\cite{brain}\cite{brainSW} \ 

The goal of this letter is to study the effect of structure on the dynamics of
sparsely connected Hopfield-type\cite{Amit}-\cite{Hopfield} attractor neural
networks. \ It is known that randomly pruning the connections of a Hopfield
net (HN) increases the storage capacity per synapse.\cite{dilution} Amongst
other questions we ask whether there is an optimal topology given a fixed
number of nodes and connections. \ The HN is of interest because it provides a
tractable toy model of collective computation and can also be viewed as an
extension of the Ising model with limited amounts of frustration and quenched
disorder.\cite{Mezard} \ We hope therefore that our results lead to further
insights into collective computation as well as ordering and disordering
processes occuring on networks. \ 

Our computations involve Hopfield nets with asynchronous updating in random
order\cite{Amit}\cite{HKP}, $p$ random stored binary pattern vectors
$\mathbf{\xi}^{\mu}$ and Hebbian\cite{Hebb} connection strengths \ %

\begin{equation}
w_{ij}=a_{ij}\sum_{\mu=1}^{p}\xi_{i}^{\mu}\xi_{j}^{\mu}\label{Hebbrule}%
\end{equation}
where $a_{ij}$ is the adjacency matrix ($a_{ij}=1$ if $i$ and $j$ are
connected, $a_{ij}=0$ otherwise). \ \ The degree of node $i$ is $k_{i}%
=\sum_{j=1}^{N}a_{ij}$. \ We always compare networks with the same number of
nodes $N=5000$ and average degree $\langle k\rangle=50$, varying only the
arrangement of connections. Each node is connected on average to 1\% of the
other nodes, \ compared to $\approx$0.1\% in the mouse
cortex.\cite{Braitenburg}\ \ \ \ \ The networks compared are: \ (1) \ a
regular one-dimensional ring of nodes, each connected to its 50 nearest
neighbors, (2) \ a random (Erd\H{o}s-Renyi\cite{E-R}) \ network, \ (3) a small
world (Watts-Strogatz\cite{WS}) net constructed from a regular lattice by
randomly rewiring local links with probability $r$, \ and (4) a scale-free
network with degree distribution $P(k)\sim k^{-3}$ with a lower cutoff of 25
generated by the Barabasi-Albert algorithm of prefential attachment\cite{BA}. \ 

\ We measured two aspects of the performance of the associative memory
networks: \ the \emph{stability} of the memorized patterns (inversely related
to the number of errors induced by crosstalk) and the nework's \emph{ability
to recognize} one of the patterns from a state with a large number of errors.
\ These two features of an associative memory are related but not identical:
\ a pattern can be stable but nonetheless have a small basin of attraction,
while on the other hand it is possible for an attractor to have a large basin
but nonetheless be imperfectly correlated with the memorized pattern. To
quantify pattern retrieval we used the overlap order parameters
\begin{equation}
m^{\mu}\equiv\frac{1}{N}\sum_{i=1}^{N}x_{i}\xi_{i}^{\mu} \label{overlap}%
\end{equation}
where $x_{i}=\pm1$ denotes the output of the $i$-th node, as well as partial
overlaps $m^{\mu}(k_{\min})=m^{\mu}(k>k_{\min})$, defined as in (\ref{overlap}%
) except that the sum runs only over those nodes whose degree exceeds
$k_{\min}$ and is normalized appropriately. \ $k_{\min}$ partitions the
network into hubs and non-hubs, and $m^{\mu}(k_{\min})$ measures recognition
of the portion of the pattern encoded in the hubs. \ The stability of the
memory patterns was measured by initializing the network to a memory state
($x_{i}=\xi_{i}^{\mu}$) \ and measuring $m_{final}^{\mu}$ after the dynamics
converged. \ The departure of $m_{final}^{\mu}$ from 1 reflects the number of
errors induced by crosstalk. \ As an indicator of the network's ability to
retrieve a pattern from a randomly corrupted version, we measured
$m_{final}^{\mu}$ when the initial overlap was $m_{init}^{\mu}=0.5$. \ \ \ We
averaged these quantities over several realizations of the topology and
patterns, varying the number of patterns $p$ to see how the performance
degrades with increasing loading. \ \ \ \
\begin{figure}
[ptb]
\begin{center}
\includegraphics[
height=2.6688in,
width=5.847in
]%
{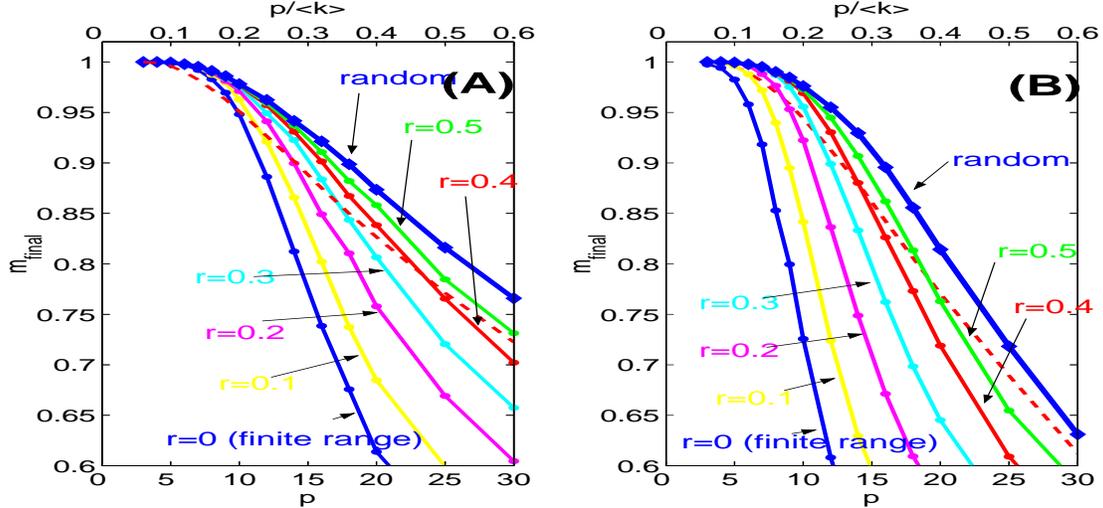}%
\caption{Performance of different networks as a function of the number of
patterns $p$ and loading ratio $p/\langle k\rangle.$ \ All have $\langle
k\rangle=50$ and $N=5000$. Solid lines: random net and small-world net ranging
from $r=0$ (i.e., a locally connected lattice) to $r=0.5$. \ Dotted line:
\ scale-free net (see also figure \ref{scalefree50}). \ Data were averaged
over patterns and network realizations, for a total of 80 trials per point.
(A) $m_{final}$ when $m_{init}=1$, a measure of the stability of the memorized
patterns against crosstalk-induced errors. \ (B) $m_{final}$ when
$m_{init}=0.5$, a measure of how successfully a pattern is retrieved from a
corrupted version. \ \ \ }%
\label{smallworld50}%
\end{center}
\end{figure}

With $\langle k\rangle=50$ and $p\lesssim50$, the networks studied are far
from the commonly studied limit where both $p$ and $N$ simultaneously approach
infinity\cite{Amit}. Therefore no discontinous overloading phase transition is
apparent, but comparisons are still possible at finite $p$ and $N$. \ \ Figure
\ref{smallworld50} shows results for the networks as a whole. \ The most rapid
degradation in both stability and retrievability occurs in the regular
lattice, and the slowest in the random net. \ Not surprisingly, the addition
of shortcuts to a regular lattice enhances pattern stability and retrieval.
\ The performance of small-world nets is intermediate between that of a
regular and a random net. \ \ The variation with rewiring probability $r$ is
not linear, however. \ A network with $r=0.5$ behaves almost as a random (
$r=1$) net. \
\begin{figure}
[ptb]
\begin{center}
\includegraphics[
height=2.4768in,
width=5.5798in
]%
{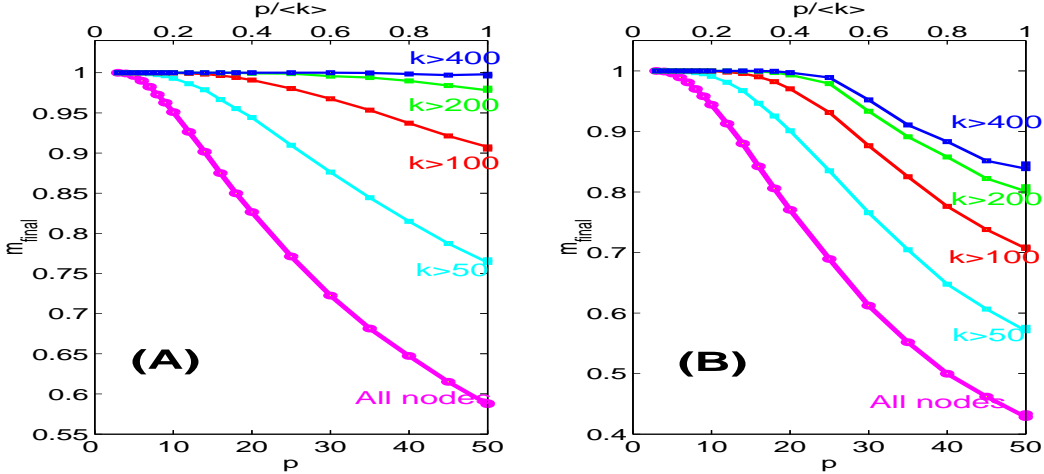}%
\caption{Performance of the whole network compared to \textquotedblleft
hub\textquotedblright\ subsets of the scale-free network, averaged over 200
trials per data point. \ (A) \ Pattern stability ( $m_{init}=1$). \ (B)
Pattern retrievability ($m_{init}=0.5$). \ \ }%
\label{scalefree50}%
\end{center}
\end{figure}

The performance of the scale-free net as a whole degrades slightly faster than
that of a random net, but the errors are not distributed evenly as is apparent
in figure \ref{scalefree50} where we examine $m^{\mu}(k_{\min})$ for $k_{\min
}=50,100,200,$ and $400$. \ \ These represent subsets having average sizes of
1235, 333, 97 and 29 nodes, respectively. \ \ From figure \ref{scalefree50}A,
it is evident that the frequency of crosstalk-induced errors decreases with
increasing degree. \ For example, the nodes with $k>k_{\min}=200$ have very
few errors even when $p=\langle k\rangle=50$. \ The nodes with $k>200$ form a
subset of approximately 100 nodes. \ A fully connected net of 100 nodes alone
would be able to store only $\ \approx14$ patterns.\cite{Amit}%
\cite{HopfieldStorage} \ Thus, the performance of the hubs in the original
network as pattern recognizers is much higher than it would be if the nodes
with $k<k_{\min}$ were pruned away. \ \ Even though the less connected nodes
are more prone to errors, they nonetheless assist the hubs in retaining the
patterns. \ \ The enhanced performance of the well-connected subset manifests
itself not only in the stability of the patterns but also in their retrieval,
as seen in figure \ref{scalefree50}B. \ The hubs are able to distinguish
clearly among a large number of patterns even if the pattern reconstruction is
incomplete (i.e. limited to the hubs.)

The lower rate of errors among the hubs is not surprising in view of the fact
that their input comes from a larger number of nodes. \ \ It can be shown
using arguments as in \cite{Amit} and \cite{HKP} that if the state of the
network is initially set to one of the patterns ($\mathbf{x}=\mathbf{\xi}%
^{\mu}$ ) then an individual node with degree $k_{i}$ experiences a
crosstalk-induced noise-to-signal ratio $(N/S)_{i}\approx\sqrt{(p-1)/k_{i}}$.
Hence the probability of a crosstalk-induced error in the $i$-th node
decreases with increasing degree. \ \ The presence of one error reduces the
strength of the signal and may increase the likelihood of additional errors,
\ resulting in a cascade of the type responsible for the abrupt overloading
phase transition that occurs in the fully connected Hopfield
network\cite{Amit}\cite{HopfieldStorage}. \ Cascades in the opposite direction
may also play a role in the reconstruction of patterns from noisy input. \ The
differences between differently connected networks thus lie not just in the
initial signal-to noise ratio but in the dynamics of the spread of error
cascades. This dynamics differs from ordinary percolation or epidemic
propagation, since the susceptibility of a node is inversely correlated with
its degree. \ In most models of epidemic propagation only one infected
neighbor suffices to infect a node, regardless of its degree. \ 

Our results are reminiscent of those for the simple Ising model: a
one-dimensional lattice is easily disordered by thermal noise\cite{Kogut}
while even a few long range connections can restore order at a finite
temperature\cite{Pekalski}, and in a scale-free network the nodes with high
degree are more strongly magnetized than those with low
degree\cite{Aleksiejuk}. \ The difference is that in the present case, \ the
disorder is induced by interference among the stored patterns and not by
stochastic noise. \ It is a quenched rather than thermal disorder. \ 

While we found that the most efficient arrangement for storage and retrieval
of patterns by the network as a whole is a random network, connections in real
brains do not appear to be fully random. \ One reason may lie in the economy
of wiring length.\cite{brain} \ The majority of connections in brains of
higher animals as well as in \textit{C. elegans} \cite{WS}\textit{ }appears to
occur between nearby neurons, while fewer paths connect more distant regions,
suggesting a small-world topology. \ Our results imply that small-world
networks with a moderate number of shortcuts can be almost as computationally
efficient as a random network while saving considerably on wiring costs. \ The
suitability of small-world networks for complex computations was also
suggested on the basis of other models.\cite{brainSW} To what extent
scale-free structures play a role in real brains remains to be seen, but our
results suggest a mechanism by which information can be centralized in the
more connected nodes while the remaining nodes, although noisy, are
nonetheless indispensable for the computation. \ It will be of interest to
study the implications of these notions for the formation of knowledge,
opinions and power structures in scientific and social networks.\cite{Opinion}

\begin{acknowledgments}
This work was supported by the NSERC of Canada. We thank H. Kwan and R. Kapral
for helpful discussions. \ 
\end{acknowledgments}

\end{document}